\documentclass{bmcart}

\usepackage[utf8]{inputenc} 
\usepackage{hyperref}
\usepackage{graphicx}

\startlocaldefs
\endlocaldefs

\begin{document}

\begin{frontmatter}

\begin{fmbox}
\dochead{Software}


\title{DeDaL: Cytoscape 3.0 app for producing and morphing data-driven and structure-driven network layouts}


\author[
   addressref={aff1,aff2,aff3},                   
   email={ulcia.liberte@gmail.com}   
]{\inits{U}\fnm{Urszula} \snm{Czerwinska}}
\author[
   addressref={aff1,aff2,aff3},
   email={Laurence.Calzone@curie.fr}
]{\inits{L}\fnm{Laurence} \snm{Calzone}}
\author[
   addressref={aff1,aff2,aff3},
   email={Emmanuel.Barillot@curie.fr}
]{\inits{E}\fnm{Emmanuel} \snm{Barillot}}
\author[
   addressref={aff1,aff2,aff3},
   corref={aff1},                       
   email={Andrei.Zinovyev@curie.fr}
]{\inits{A}\fnm{Andrei} \snm{Zinovyev}}


\address[id=aff1]{
  \orgname{Institut Curie}, 
  \street{26 rue d'Ulm},                     %
  \city{Paris},                              
  \cny{FR}                                    
}
\address[id=aff2]{%
  \orgname{INSERM U900},
  \city{Paris},
  \cny{FR}
}
\address[id=aff3]{%
  \orgname{Mines Paris Tech},
  \city{Fontainebleau},
  \cny{FR}
}



\end{fmbox}


\begin{abstractbox}

\begin{abstract} 
{\bf Background:} Visualization and analysis of molecular profiling data together with biological networks
are able to provide new mechanistical insights into biological functions.
Currently, high-throughput data are usually visualized
on top of predefined network layouts which are not always adapted to a given data analysis task.
We developed a Cytoscape app which allows to construct biological network layouts based
on the data from molecular profiles imported as values of nodes attributes.

{\bf Results:} DeDaL is a Cytoscape 3.0 app which uses linear and non-linear algorithms of dimension reduction to produce
data-driven network layouts based on multidimensional data (typically gene expression).
DeDaL implements several data pre-processing and layout
post-processing steps such as continuous
morphing between two arbitrary network layouts and aligning one
network layout with respect to another one by rotating and mirroring.
Combining these possibilities facilitates creating insightful
network layouts representing both structural network features
and the correlation patterns in multivariate data.

{\bf Conclusions:} DeDaL is the first method allowing to construct biological network layouts from high-throughput data.
DeDaL is freely available for downloading together with step-by-step tutorial at http://bioinfo-out.curie.fr/projects/dedal/.

\end{abstract}




\end{abstractbox}
%

\end{frontmatter}



\section*{Background}
One of the major challenges in systems biology is to combine in a meaningful way the large corpus of knowledge in molecular biology recapitulated in the form of large interaction networks together with high-throughput omics data produced at increasing rate, in order to advance our understanding of biology or pathology \cite{Barillot2012}.

There exists numerous methods using biological networks for making insightful high-throughput data analysis \cite{Barillot2012}.
One can distinguish three large groups of such methods: (1) mapping the data on top of a pre-defined biological network layout,
(2) identification of subnetworks in a global network possessing certain properties computed from the data (such as subnetworks enriched
with differentially expressed genes), (3) using biological network structure for pre-processing the high-throughput data (for example,
for ``smoothing" the discrete mutation data).

Quantitative omics data can be mapped on top of a pre-defined biological network layout. Currently, most of the pathway
databases (such as KEGG \cite{Kanehisa2012}, Reactome \cite{Croft2011}) provides such a possibility, using simple data visualization tools.
Omics data visualization tools using networks are constantly improved and become more elaborated \cite{Gehlenborg2010}.
For example, VANTED tool \cite{Klukas2010} creates a classification tree according to the KEGG pathway hierarchy
and shows a biological network with omics data as barplots
or pie-charts attached to the nodes which allows to visualize more complex data than by
simple node coloring. NaviCell \cite{Kuperstein2013} and related pathway
database Atlas of Cancer Signalling Network (ACSN) together with standard heat maps and barplots provide
more flexible data visualization tools such as glyphs (symbols with configurable shape, size and color) and map staining (using the network background for visualization)
\cite{Kuperstein2015}. An interesting approach for data visualization using biological networks was developed in NetGestalt online tool
\cite{Shi2013} which uses a NetSAM R package to create modules by hierarchical ordering of the network in one
dimension and visualizes high-throughput data accordingly to a chosen track
as a combination of barplots and heat maps.

Omics data are used to identify overexpressed or enriched subnetworks.
For example, in \cite{Ulitsky2007} expression data were combined with network information in
order to identify under- or overexpressed subnetworks in Huntington`s disease and
breast cancer. Inspired by this method, several
Cytoscape plug-ins were developed and applied to various omics data in order to find
connected sub-components where most of the genes are differentially expressed or co-expressed \cite{Cline2007,Alcaraz2012}.
A recent review on integrating molecular profiles with networks in order to find ``network modules"
can be found in \cite{Mitra2013}.

Using projection of the high-throughput data into the basis of functions smooth on a biological network graph was suggested
in \cite{Rapaport2007}. Recently, biological networks were used to regularize the genome-wide mutational landscapes (which are sparse) in cancer, using
network smoothing methods \cite{Hofree2013}.

However, none of the methods cited above had a purpose to visualize high-throughput data by
computing a specific network layout based on the omics data themselves, which would combine both network structure
and the data from the network node attributes. Some of the existing Cytoscape layout algorithms (such as Group Attributes Layout)
allow using the values of single node attributes, but this possibility is currently under-developed.
We believe that using networks for visualizing and analyzing data requires methods that would
be able to create more suitable and adapted for a particular task biological network layouts.

Mathematically speaking, molecular entities exist    in two metric spaces: in space of biological functions, where the distance between two molecules can be defined by the number of steps (edges) in a graph defining pairwise functional relations (such as protein-protein interactions) along the shortest path connecting them; and in data space, where the distance between two molecules is defined by the proximity of the corresponding numerical descriptors (such as expression profiles). The network distances are usually visualized by designing a 2D or 3D layout, representing the network structure. Visualization of distances in data space is achieved by data dimension reduction methods (such as PCA) projecting multidimensional vectors in 2D or 3D space. No methods were developed so far for performing dimension reduction in Cytoscape and mixing the two types of visual representations together. Having this in mind, we've developed DeDaL, a Cytoscape 3.0 app for mixing purely data-driven and purely structure-driven network layouts.

\section*{Implementation}

DeDaL is a {\it simplified} Cytoscape 3 app implemented in Java language. For computing linear and non-linear principal manifolds, DeDaL uses VDAOEngine Java library, developed by AZ (http://bioinfo-out.curie.fr/projects/elmap/). For computing the eigenvectors of a symmetric Laplacian matrix, Colt library has been used (http://acs.lbl.gov/ACSSoftware/colt/). Internal graph implementation is re-used from BiNoM Cytoscape plugin \cite{Zinovyev2008, Bonnet2013, Bonnet2013a}. The source code of DeDaL is available at http://bioinfo-out.curie.fr/projects/dedal.

\subsection*{Producing data-driven network layouts}

Data-driven network layout (DDL) is produced by DeDaL by positioning the nodes of the network according to their projection from the multidimensional data space of associated numerical vectors into some 2D space. DeDaL implements three algorithms for performing this dimension reduction: (1) Projection onto a plane of two selected principal components; (2) Projection onto a non-linear 2D surface approximating the multidimensional data distribution, i.e. principal manifold, computed by the method of elastic maps \cite{gorban2001method, gorban2005elastic, Gorban2008Principal, Gorban2010}; (3) Using (1) or (2) preceded by network-based regularization (smoothing) of the data, based on computing the $k$ first eigen vectors of the Laplacian matrix of the network graph and projecting data into this subspace (as suggested in \cite{Rapaport2007}).

DeDaL implements specific data pre-processing and resulting layout post-processing steps. Pre-processing steps include (1) selecting only nodes whose associated numerical vectors (imported as tables to Cytoscape) are sufficiently complete and (2) optional double centering of the data matrix. Post-processing of the resulting layout includes (1) avoiding overlap between node positions by moving them in a random direction at a small distance; (2) moving the outliers (nodes positioned too distantly from other nodes) closer to the barycenter of the data distribution; (3) placing the nodes with missing data into the mean point of the position of their network neighbours.

In the future we will exploit a possibility to project the data into 3D and will implement additional dimension reduction algorithms such as multidimensional scaling.

\subsection*{Manipulating network layouts in DeDaL}

In order to allow comparing the resulting DDLs with standard layouts produced by Cytoscape and transform one into another, DeDaL implements simple layout morphing and aligning methods. Morphing of two network layouts is performed by a linear transformation, moving matched nodes along straight lines. DeDaL provides a convenient user dialog for morphing one layout into another in which a user can use slider and immediately appreciate the morphing result. The morphing operation provides poor results if one layout is systematically rotated or flipped with respect to the node positions in another one. DeDaL allows aligning two network layouts by rotating and mirroring, and minimizing the Euclidean distance between two layouts.

\subsection*{Double-centering the data matrix}

The data matrix is optionally double-centered by subtracting from each matrix entry the mean value calculated over the corresponding matrix row and the mean value calculated over the matrix column, and by adding the global mean value computed over all matrix entries. This procedure allows to eliminate some global biases in the data such as the global differences in average fluorescence intensity of different probes in microarray data.

\subsection*{Network-based smoothing of data}

Network data smoothing is made in DeDaL as it was suggested in \cite{Rapaport2007}. For a graph representing the biological network, its Laplacian and all its eigenvectors are computed. These vectors define a new orthonormal basis in the multidimensional data space. To smooth the values of the data matrix, the initial multidimensional vector associated to a datapoint is projected into the subspace spanned by the first $k$ eigenvectors of the graph's Laplacian. DeDaL smoothing parameter is the  $p_{ns}=1-\frac{k-(n_c+2)}{N-(n_c+2)}, p_{ns}\in [0;1]$, where $n_c$ is the number of connected components in the graph and $N$ is the number of nodes on the graph. Therefore, $p_{ns}=0$ corresponds to $k=N$, i.e. when no smoothing is performed and all eigenvectors are used, while $p_{ns}=1$ corresponds to $k=(n_c+2)$ and first two non-degenerated eigenvectors are used to smooth the data (the data become effectively three-dimensional, with the first dimension corresponding to the average value of the data matrix computed over each connected component of the graph).

\subsection*{Exporting the pre-processed data}

The results of pre-processing the data for a given network can be exported to a file. Actually, two files are created: one in a simple tab-delimited format suitable for further analysis in most statistical software packages and another file in the ``.dat" format, suitable for analysis in ViDaExpert multidimensional data visualization tool \cite{GorbanVidaExpert2014}. This possibility can be used, for example, for network smoothing of an expression dataset for further application in any machine learning algorithm (clustering, classification). For this purpose, DeDaL can be also used in a command line mode (see examples at the web-site).

\subsection*{Computing principal components}

The principal components in DeDaL are computed using singular value decomposition, computed by the method allowing to use missing data values without pre-imputing them, as it is described in \cite{Gorban2009}. Data points, containing more than 20\% of missing values are filtered out from the analysis. DeDaL computes the 10 first principal components if there is more than 10 data points, and $k$ principal components if there is $k+1$ data points, $k<10$. After computing the principal components, DeDaL reports the amount of variance explained by each of the principal components.

\subsection*{Continuous layout morphing}

Morphing two network layouts is performed by a simple linear transformation. A node having position $(x_{11},x_{12})$ in the initial layout and the position $(x_{21},x_{22})$ in the target layout is placed during the morphing procedure in the position $(p\times x_{21}+(1-p){x_{11}}, p\times x_{22}+(1-p)x_{12})$, where $p\in [0;1]$ is the morphing parameter representing the fracture of distance between the initial and target node positions along the straight line.

\subsection*{Aligning two network layouts by rotation and mirroring}

Morphing between two network layouts might be meaningless if all nodes in one layout are systematically rotated or flipped with respect to the node positions in another layout. This situation is often the case when producing the pure data-driven layout and comparing it to the initial structure-driven layout. In this case, DeDaL allows minimizing the Euclidean distance between two layouts defined as the sum of squared Euclidean distances between all matched nodes with respect to all possible rotations and mirroring of one of the layouts. To do this, a user should simply check the corresponding checkbox in the user dialog before starting to apply layout morphing. Also, a user can align several network layouts to one chosen reference network layout, using a separate "Layout aligning" dialog. For example, it is usually useful to align the structure-driven layouts to the PCA-based data-driven layout.

\section*{Results}

\subsection*{Using The Cancer Genome Atlas transcriptome data and Human Protein Reference Database network}

We used The Cancer Genome Atlas (TCGA) transcriptomic dataset for breast cancer (548 patients)\cite{TCGA_BRCA2012} and Human Reference Protein Database (HRPD) database \cite{Peri2004} as a source of protein-protein interaction network.

Firstly, as an example of a small subnetwork, we selected proteins involved in Fanconi DNA repair pathway \cite{Moldovan2009a} as it is defined in Atlas of Cancer Signaling Network (ACSN, \url{http://acsn.curie.fr}). For node coloring, we mapped the value of the t-test computed for the gene expression difference between the basal-like (one of the molecular subtypes of breast cancer, significantly contributing to the intertumoral variability) and non basal-like breast tumours. We've imported the TCGA data in Cytoscape and applied DeDaL for the transcription levels of the genes in the subnetwork (Figure 1).

\begin{figure}[h!]
\begin{center}
\includegraphics[width=120mm]{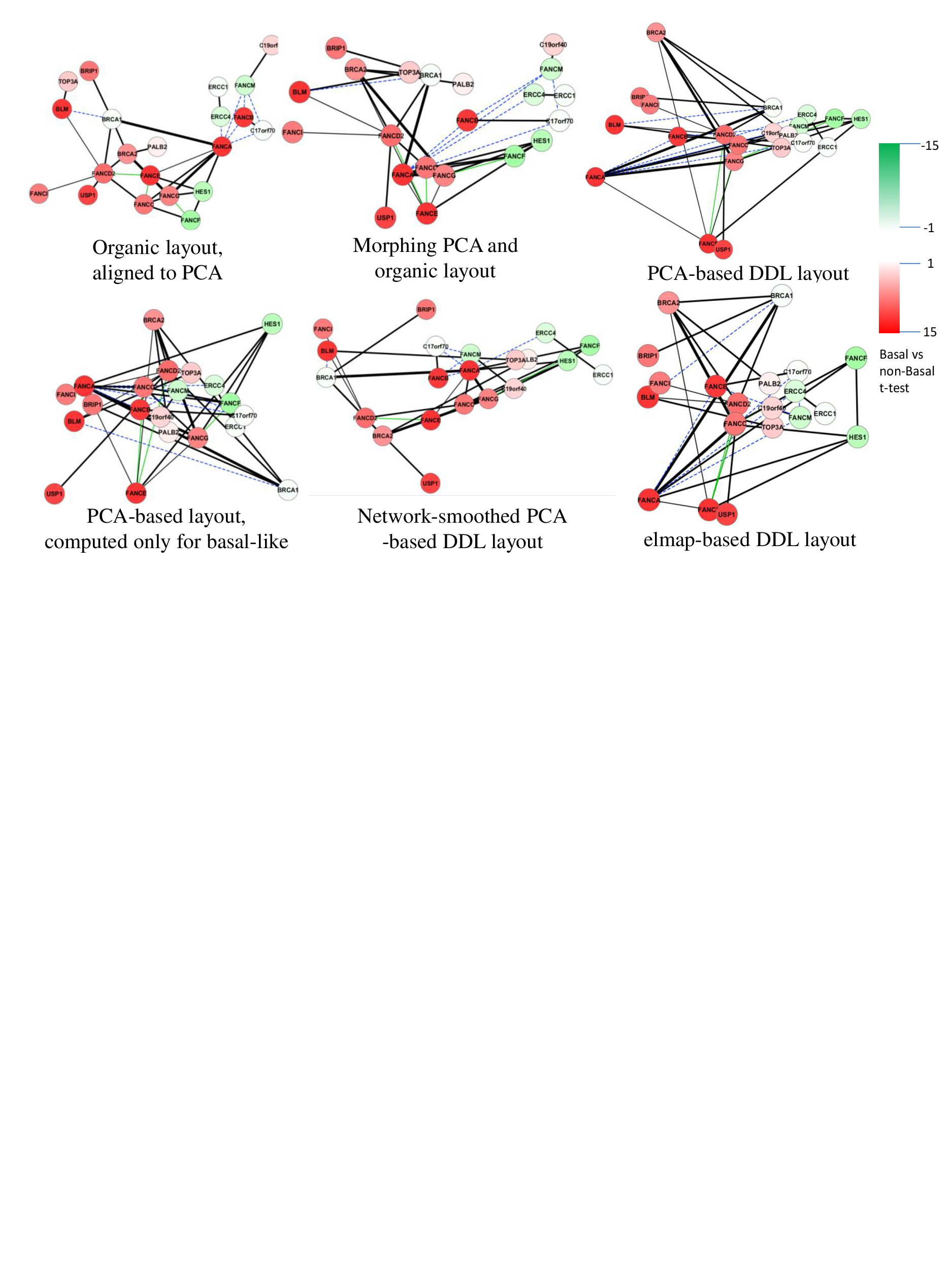}
\caption{\csentence{Using DeDaL for visualizing Fanconi pathway in breast cancer} Top row from left to right: Standard organic layout, PCA-based DDL, morphing two previous layouts at half-distance. Bottom row from left to right: PCA-based DDL computed only for basal-like tumours (note change in position of BRCA1 gene), PCA applied to network-smoothed profile,
DDL computed using elastic map (elmap) algorithm for computing non-linear principal manifold.}
\label{Figure_DeDaL}
\end{center}
\end{figure}

One can see (Figure~\ref{Figure_DeDaL}, top right) that the first principal component sorts the nodes accordingly to the t-test, because in this case the first principal component is associated with the basal-like breast cancer subtype. The second principal component gives additional information such as that the expression levels of BRCA2 and FANCE are differently modulated though both are upregulated in the basal-like subtype. Morphing the organic network layout with the PCA-based layout moves position of some of the genes, keeping the general pattern of PCA preserved, while better reflecting the network structure.

We've also applied PCA-based DDL to the subset of basal-like breast tumours (Figure~\ref{Figure_DeDaL}, bottom left) which showed the specific role of BRCA1 gene in this subtype (which is known). Also, the position of USP1 gene has significantly changed with respect to the PCA-based DDL produced for the whole set of samples. This demonstrates the ability of DeDaL to produce network layouts specific for a particular cancer subtype.

Application of network smoothing is demonstrated at Figure~\ref{Figure_DeDaL}, bottom middle. The layout preserves the general pattern of the PCA-based DDL, while better visualizing the network structure, and moving some proteins into a different position. For example, BRCA1 gene is moved to left because it is connected to several genes overexpressed in basal-like breast cancer subtype. Figure~\ref{Figure_DeDaL}, bottom right, shows application of non-linear PCA to data dimension reduction. This network layout better resolves the relations between some gene expression levels such as FANCF and HES1 and the roles of BRCA1 and BRCA2 in Fanconi DNA repair pathway.

\begin{figure}[h!]
\begin{center}
\includegraphics[width=120mm]{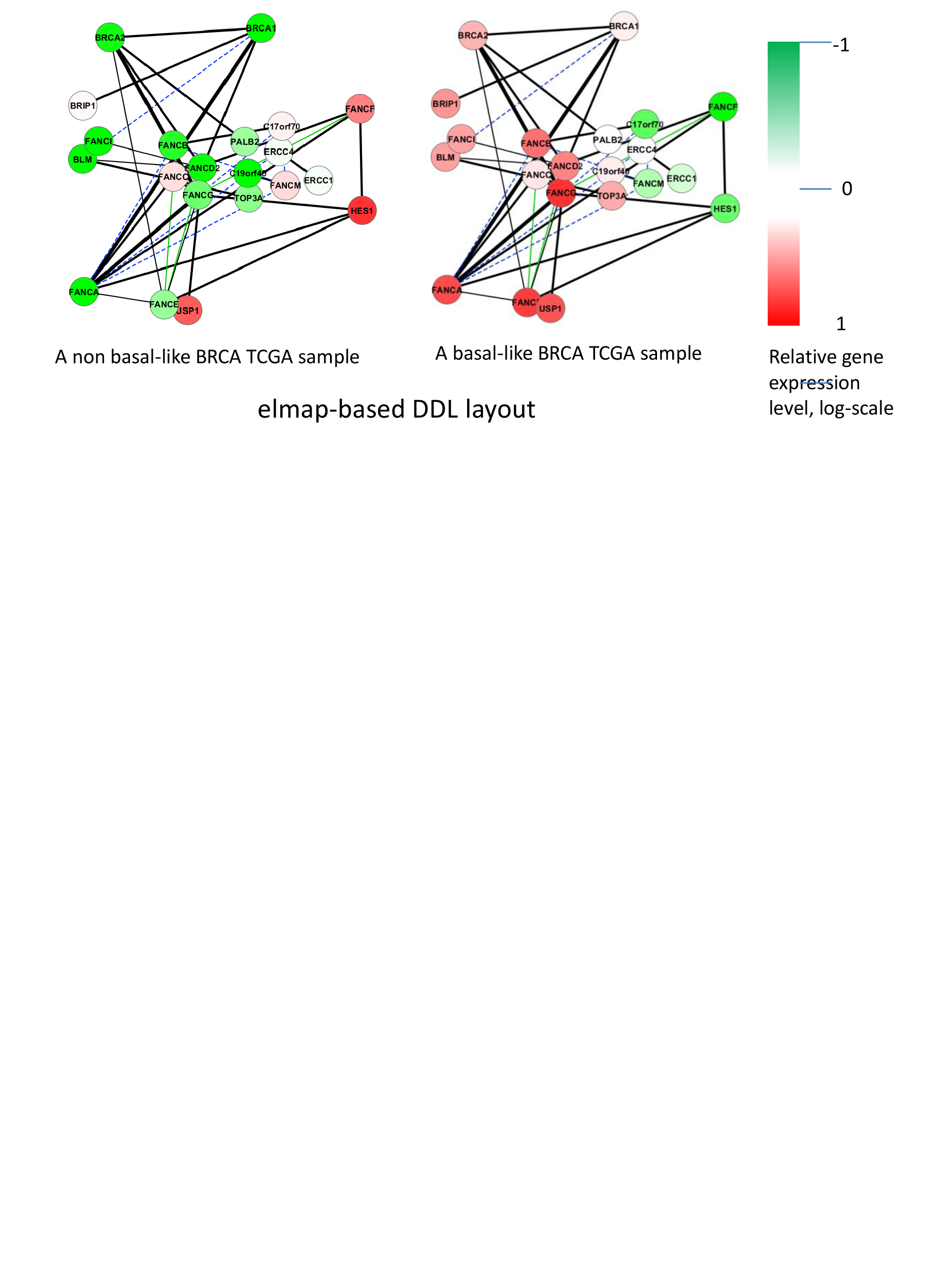}
\caption{\csentence{Using DeDaL for showing individual sample gene expression profiles}. {Expression profiles on the Fanconi pathway genes for two randomly chosen samples (one basal-like and one non basal-like) from TCGA breast cancer cohorts are shown. The expression levels are computed as relative to the mean value over the whole cohort.}}
\label{Fanconi_BAS100_NBAS187}
\end{center}
\end{figure}

DDLs produced by DeDaL can serve to better visualize expression pattern in individual samples. Examples of using elastic map (elmap)-based DDL for distinguishing one randomly chosen basal-like and one non basal-like expression profiles of Fanconi pathway is shown in Figure~\ref{Fanconi_BAS100_NBAS187}. Unlike organic layout, DDL allows quickly evaluate the general trend of the expression profile and detect exceptions from this trend like USP1 gene, known to be a biomarker of genomic instability and Fanconi anemia phenotype \cite{Kim2009}, and overexpressed in both samples.

\begin{figure}[h!]
\begin{center}
\includegraphics[width=120mm]{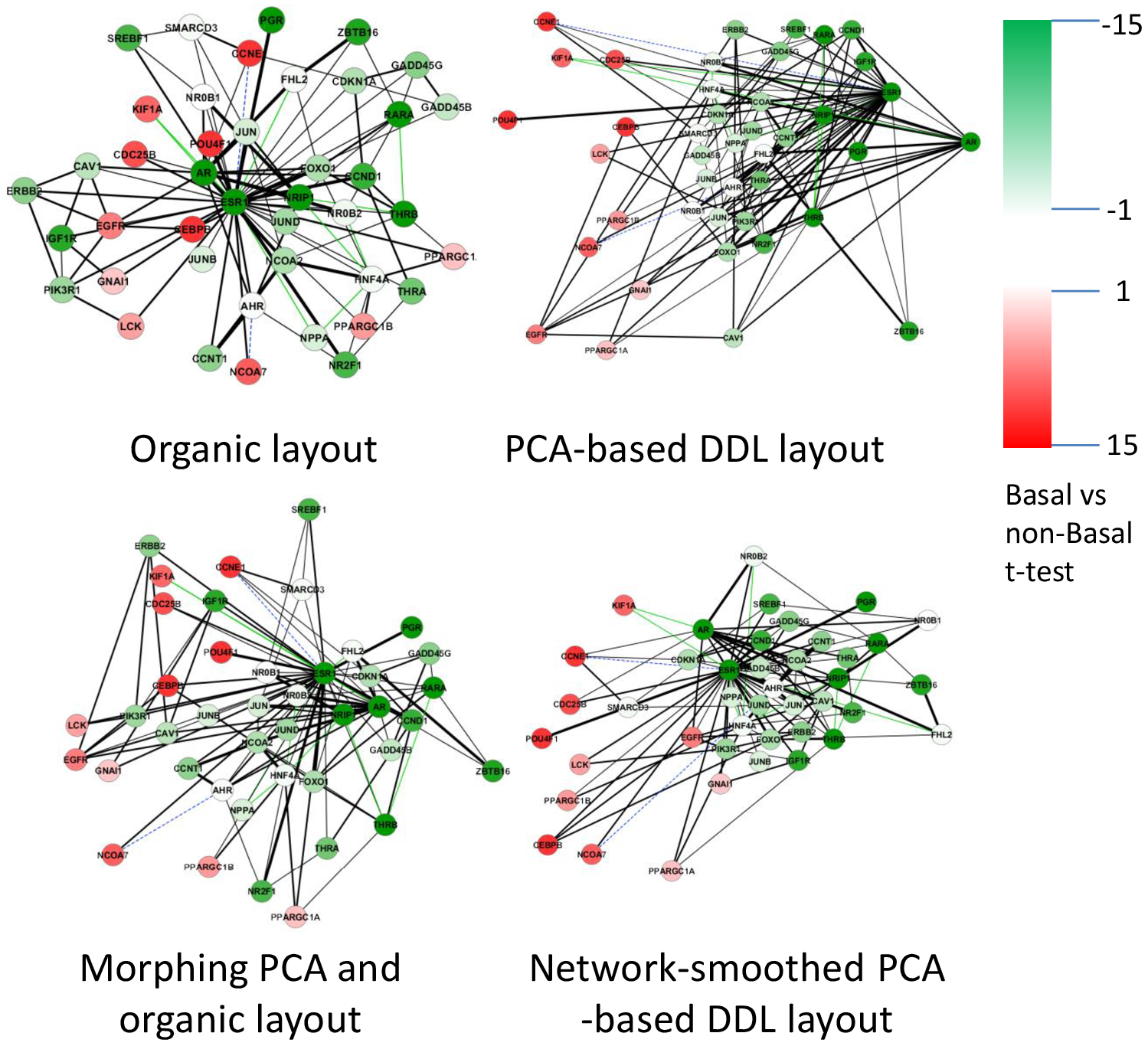}
\caption{\csentence{Using DeDaL for visualizing network of genes interacting with ESR1}. {DeDaL allows mixing purely structure-driven network layout (top left) with purely data-driven network layout (top right) by morphing them (bottom left, which is the half-distance between two upper layouts). Bottom right is the same as PCA-based layout (top right) but network smoothing was performed before applying PCA.}}
\label{ESR1_DeDaL}
\end{center}
\end{figure}

Secondly, we selected all proteins interacting with ESR1 protein (Figure~\ref{ESR1_DeDaL}). In this case, the second principal component shows, for example, that the expression levels of EGFR and CCNE1 are differently modulated though both are upregulated in the basal-like subtype. PCA layout also highlights a particular pattern of expression of some hub genes such as AR or EGFR, and shows that underexpressed genes in basal-like subtype forms more tightly connected subnetwork. Morphing the original organic network layout with the PCA-based layout moves position of some of the proteins, keeping the general pattern of PCA preserved. For example, underexpressed PIK3R1, IGFR1 and ERBB2 genes are moved on the left because each of them is connected to several overexpressed genes. Application of network smoothing drives the hub genes to the center of the layout, because of averaging over the hub's neighbors. It produces more regular pattern of network connections but approximately conserves the neighborhood relations in PCA layout. Therefore, combining DeDaL methods allows different ways of mixing network structure and high-throughput data for producing new network layouts.

\subsection*{Visualizing genetic interactions}

Genetic interactions between two genes happen in the case where their functions are synergistic (negative interactions) or mutually alleviating (positive interactions). The strength of genetic interactions is characterized by an epistatic score which quantifies deviation from a simple multiplicative model. In the global network of genetic interactions, each gene can be characterized by its epistatic profile, which is a vector of epistatic scores with all other genes \cite{Costanzo2010}. It is shown that the genes with similar epistatic profiles tend to have similar cellular functions.

\begin{figure}[h!]
\begin{center}
\includegraphics[width=120mm]{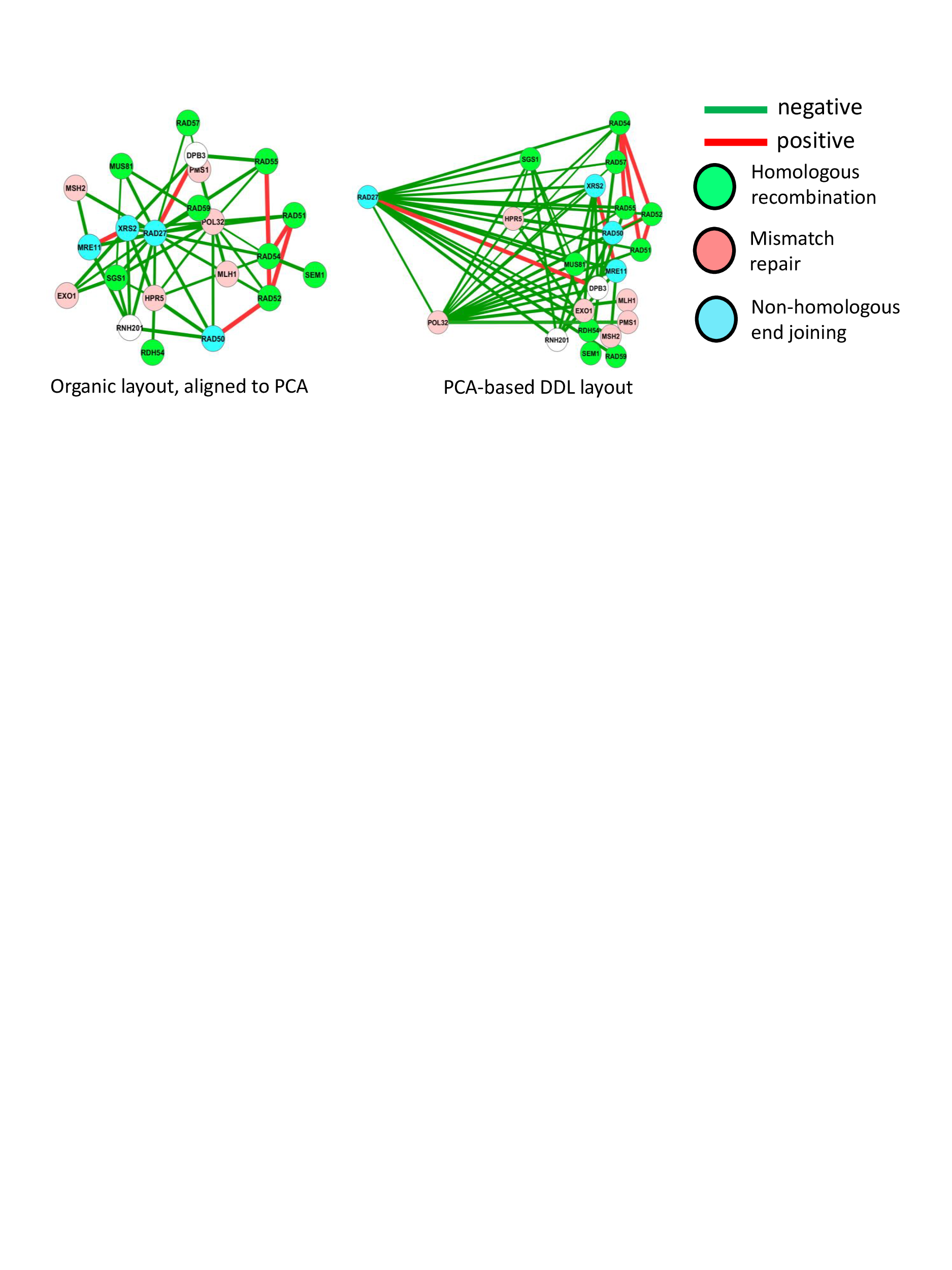}
\caption{\csentence{Using DeDaL for visualizing network of genetic interactions between yeast genes involved in DNA repair}. {Red and green edges denote positive and negative genetic interactions correspondingly. Different node colors indicate three distinct DNA repair pathways in yeast.}}
\label{Genetic_Interactions}
\end{center}
\end{figure}

We applied DeDaL to create a DDL layout for a group of yeast genes involved in DNA repair and replication. The genetic interactions between these genes and the epistatic profiles (computed only with respect to this group of genes) were used from \cite{Costanzo2010}. The definitions of DNA repair pathways were taken from KEGG database \cite{Kanehisa2012}.
Figure~\ref{Genetic_Interactions} shows the difference between application of the standard organic layout for this small network of genetic interactions and PCA-based DDL (computed here without applying data matrix double-centering to take into account tendencies of genes to interact with smaller or larger number of other genes). PCA-based DDL in this case groups the genes with respect to their epistatic profiles. Firstly, local hub genes RAD27 and POL32 have distinct position in this layout. Secondly, PCA-based DDL roughly groups the genes accordingly to the DNA repair pathway in which they are involved. For example, it shows that Non-homologous end joining DNA repair pathway is closer to Homologous recombination (HR) pathway than to the Mismatch repair pathway. It also underlines that some homologous recombination genes (such as RDH54) are characterized by a different pattern of genetic interactions than the ``core" HR genes RAD51, RAD52, RAD54, RAD55, RAD57,

\subsection*{Visualizing attractors of a Boolean model}

\begin{figure}[h!]
\begin{center}
\includegraphics[width=120mm]{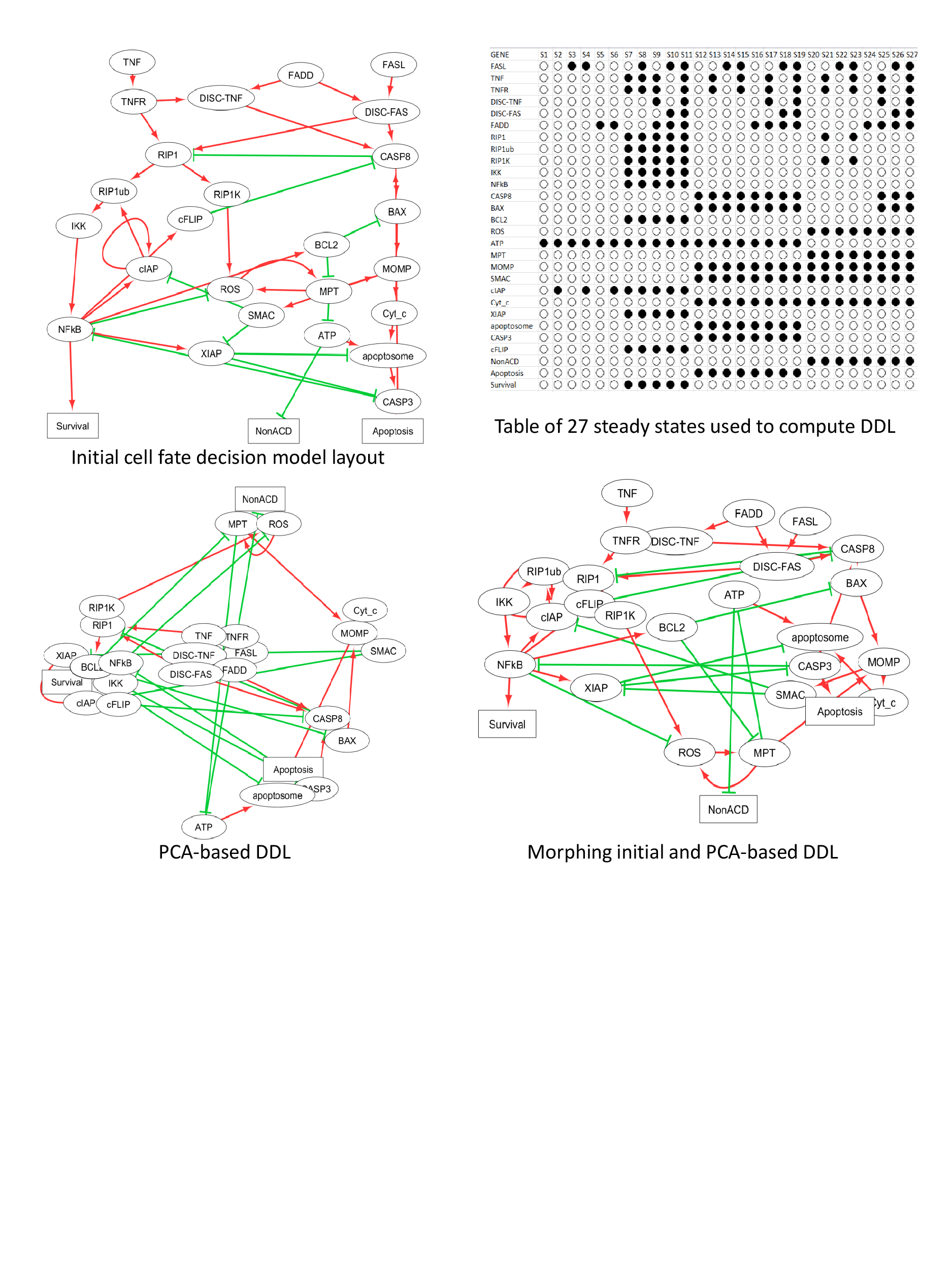}
\caption{\csentence{Using DeDaL for visualizing results of a Boolean model simulation}. {Table of computed steady states is used to group the nodes with similar states in similar conditions (shown in top right corner). In the influence diagram green edges signify inhibitory and red edges - activating relations. }}
\label{Model1}
\end{center}
\end{figure}

\begin{figure}[h!]
\begin{center}
\includegraphics[width=120mm]{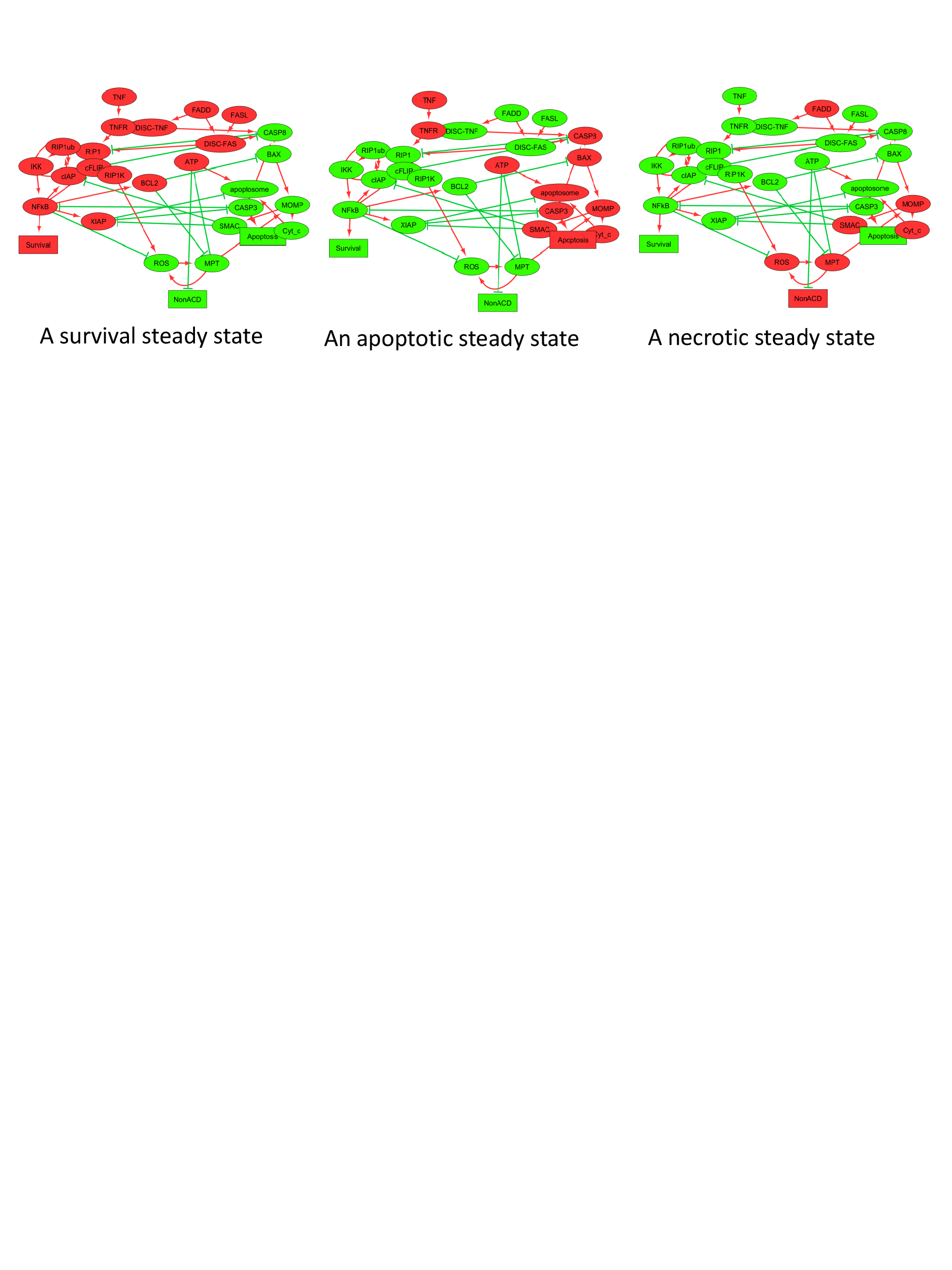}
\caption{\csentence{Using DeDaL for visualizing results of a Boolean model simulation}. {Visualization of three steady states of the model, with green and red denoting inactive (FALSE) and active (TRUE) states of the node correspondingly.}}
\label{Model2}
\end{center}
\end{figure}

In this example we used the Boolean model of cell fate decisions between survival, apoptosis and non-apoptotic cell death (such as necrosis) published in \cite{Calzone2010}, to group the nodes of the influence diagram accordingly to their co-activation patterns in the logical steady states. The table of steady states was taken from \cite{Calzone2010} (Figure~\ref{Model1}, top right) and used to compute the PCA-based DDL (Figure~\ref{Model1}, bottom left). In this DDL, nodes in close positions have similar pattern of activation in steady states (such as RIP1 and RIP1K). We used morphing PCA-based DDL and the initial layout of the model (as it was designed in \cite{Calzone2010}) to visualize several stable  states corresponding to different cell fates (Figure~\ref{Model2}). In this layout co-activated nodes tend to form compact groups. Therefore, DeDaL can be used to design layouts of mathematical models of biological networks, using the solutions of the model.

\section*{Conclusions}

DeDaL Cytoscape plugin combines the classical and advanced data dimension reduction methods with the algorithms of network layouting inside Cytoscape environment. This ability can be used in a number of ways and applications, some of them are suggested in this manuscript.

The application of DeDaL is not limited to producing data-driven network layouts. More generally, DeDaL allows applying dimension reduction of the multivariate data associated with the nodes of any Cytoscape network, optionally using the structure of the network, and export the results for further analysis by any suitable algorithms.

In the future works an effort will be made to project the data in the three dimensional space. The user will be able to rotate freely in all three dimensions and better see patterns which are difficult to represent in 2D space. The software will be also completed with alternative dimension reduction algorithms such as multidimensional scaling which will extend the data modeling possibilities, better answering to specific user's needs.


\section*{Availability and requirements}

{\bf Project name:} DeDaL: Data-Driven Network Layouting

{\bf Project home page:} http://bioinfo-out.curie.fr/projects/dedal/

{\bf Operating system(s):} Platform independent

{\bf Programming language:} Java

{\bf Other requirements:} Java 1.6 or higher, Cytoscape 3.0 or higher

{\bf License:} GNU GPL

{\bf Any restrictions to use by non-academics:} free for any non-commercial use

\begin{backmatter}

\section*{Competing interests}
  The authors declare that they have no competing interests.

\section*{Author's contributions}
      AZ and UC developed the algorithm. UC and AZ wrote the application code. UC, LC and AZ provided the examples of using DeDaL. AZ and UC wrote the manuscript. AZ, LC, EB have read and revised the manuscript.

\section*{Acknowledgements}
    We acknowledge Eric Viara and Eric Bonnet for their help in implementing DeDaL and Loredana Martignetti for helping analysing the data. All authors are members of the team ``Computational Systems Biology of Cancer". The work is supported by ITMO Cancer SysBio program, (INVADE project) and, the grant "Projet Incitatif et Collaboratif: Computational Systems Biology Approach for Cancer" from Institut Curie and by Institut National de la Sant\'e et de la Recherche M\'edicale (U900 budget).


\bibliographystyle{bmc-mathphys} 
\bibliography{Czerwinska_etal_BMCSysBiol}      

\end{backmatter}
\end{document}